\documentclass[conference]{IEEEtran}
\IEEEoverridecommandlockouts
\usepackage[noadjust]{cite}

\usepackage{amsmath,amssymb,amsfonts}
\usepackage{algorithmic}
\usepackage{graphicx}
\usepackage{textcomp}
\usepackage{xcolor}

\usepackage{breakurl}
\usepackage{subcaption}
\usepackage{dblfloatfix}

\usepackage{cite}
\usepackage{amsmath,amssymb,amsfonts}
\usepackage{algorithmic}
\usepackage{graphicx}
\usepackage{textcomp}

\usepackage{array,graphicx}
\usepackage{booktabs}
\usepackage{pifont}
\usepackage{helvet}  
\usepackage{courier}  
\usepackage[hyphens,spaces,obeyspaces]{url}

\usepackage{graphicx}  

\usepackage{dblfloatfix}

\usepackage{tabu}
\usepackage{lscape}
\usepackage{longtable}
\usepackage{multirow}
\usepackage{graphicx}
\usepackage{tabularx}

\def\BibTeX{{\rm B\kern-.05em{\sc i\kern-.025em b}\kern-.08em
    T\kern-.1667em\lower.7ex\hbox{E}\kern-.125emX}}
\begin{document}

\title{Understanding the Radical Mind: Identifying Signals to Detect Extremist Content on Twitter}

\author{\IEEEauthorblockN{Mariam Nouh\IEEEauthorrefmark{1},
Jason R.C. Nurse\IEEEauthorrefmark{2}, and Michael Goldsmith\IEEEauthorrefmark{1}}
\IEEEauthorblockA{\IEEEauthorrefmark{1}Department of Computer Science, University of Oxford, UK\\
Email: \{mariam.nouh, michael.goldsmith\}@cs.ox.ac.uk}
\IEEEauthorblockA{\IEEEauthorrefmark{2}School of Computing, University of Kent, UK\\ Email: j.r.c.nurse@kent.ac.uk}}

\IEEEoverridecommandlockouts
\IEEEpubid{\begin{minipage}{\textwidth}\ \\[12pt]
	\copyright 2019 IEEE. Personal use of this material is permitted. Permission from\\ IEEE must be obtained for all other uses, in any current or future media,\\ including reprinting/republishing this material for advertising or promotional\\purposes,creating new collective works, for resale or redistribution to servers\\ or lists, or reuse of any copyrighted component of this work in other works.
\end{minipage}} 

\maketitle

\IEEEpubidadjcol

\begin{abstract}
The Internet and, in particular, Online Social Networks have changed the way that terrorist and extremist groups can influence and radicalise individuals. Recent reports show that the mode of operation of these groups starts by exposing a wide audience to extremist material online, before migrating them to less open online platforms for further radicalization. Thus, identifying radical content online is crucial to limit the reach and spread of the extremist narrative. In this paper, our aim is to identify measures to automatically detect radical content in social media. We identify several signals, including textual, psychological and behavioural, that together allow for the classification of radical messages. Our contribution is three-fold: (1) we analyze propaganda material published by extremist groups and create a contextual text-based model of radical content, (2) we build a model of psychological properties inferred from these material, and (3) we evaluate these models on Twitter to determine the extent to which it is possible to automatically identify online radical tweets. Our results show that radical users do exhibit distinguishable textual, psychological, and behavioural properties. We find that the psychological properties are among the most distinguishing features. Additionally, our results show that textual models using vector embedding features significantly improves the detection over TF-IDF features. We validate our approach on two experiments achieving high accuracy. Our findings can be utilized as signals for detecting online radicalization activities.

\end{abstract}


\section{Introduction}\label{sec:introduction} 

The rise of Online Social Networks (OSN) has facilitated a wide application of its data as sensors for information to solve different problems. For example, Twitter data has been used for predicting election results, detecting the spread of flu epidemics, and a source for finding eye-witnesses during criminal incidents and crises~\cite{morstatter2014finding},~\cite{tumasjan2010predicting}. This phenomenon is possible due to the great overlap between our online and offline worlds. Such seamless shift between both worlds has also affected the modus operandi of cyber-criminals and extremist groups~\cite{Edwards2013}. They have benefited tremendously from the Internet and OSN platforms as it provides them with opportunities to spread their propaganda, widen their reach for victims, and facilitate potential recruitment opportunities. For instance, recent studies show that the Internet and social media played an important role in the increased amount of violent, right-wing extremism~\cite{NYT}. Similarly, radical groups such as Al-Qaeda and ISIS have used social media to spread their propaganda and promoted their digital magazine, which inspired the Boston Marathon bombers in 2010~\cite{rand}.

To limit the reach of cyber-terrorists, several private and governmental organizations are policing online content and utilising big data technologies to minimize the damage and counter the spread of such information. For example, the UK launched a Counter Terrorism Internet Referral Unit in 2010 aiming to remove unlawful Internet content and it supports the police in investigating terrorist and radicalizing activities online. The Unit reports that among the most frequently referred links were those coming from several OSNs, such as Facebook and Twitter~\cite{Edwards2013}. Similarly, several OSNs are constantly working on detecting and removing users promoting extremist content. In 2018, Twitter announced that over $1.2$ million accounts were suspended for terrorist content~\cite{TwitterPubliPolicy}. 

Realizing the danger of violent extremism and radicalization and how it is becoming a major challenge to societies worldwide, many researchers have attempted to study the behaviour of pro-extremist users online. Looking at existing literature, we find that a number of existing studies incorporate methods to identify distinguishing properties that can aid in automatic detection of these users~\cite{Ashcroft2015,rowe2016}. However, many of them depend on performing a keyword-based textual analysis which, if used alone, may have several shortcomings, such as producing a large number of false positives and having a high dependency on the data being studied. In addition, it can be evaded using automated tools to adjust the writing style.

Another angle for analyzing written text is by looking at the psychological properties that can be inferred regarding their authors. This is typically called psycholinguistics, where one examines how the use of the language can be indicative of different psychological states. Examples of such psychological properties include introversion, extroversion, sensitivity, and emotions. 
One of the tools that automates the process of extracting psychological meaning from text is the Linguistic Inquiry and Word Count (LIWC)~\cite{liwc} tool. This approach has been used in the literature to study the behaviour of different groups and to predict their psychological states, such as predicting depression~\cite{de2013predicting}. More recently, it has also been applied to uncover different psychological properties of extremist groups and understand their intentions behind the recruitment campaigns~\cite{vergani2017language}. 
 
Building on the findings of previous research efforts, this paper aims to study the effects of using new textual and psycholinguistic signals to detect extremist content online. These signals are developed based on insights gathered from analyzing propaganda material published by known extremist groups. In this study, we focus mainly on the ISIS group as they are one of the leading terrorist groups that utilise social media to share their propaganda and recruit individuals. We analyze the propaganda material they publish in their online English magazine called ~\textit{Dabiq}, and use data-mining techniques to computationally uncover contextual text and psychological properties associated with these groups. From our analysis of these texts, we are able to extract a set of signals that provide some insight into the mindset of the radical group. 
This allows us to create a general radical profile that we apply as a signal to detect pro-ISIS supporters on Twitter. Our results show that these identified signals are indeed critical to help improve existing efforts to detect online radicalization.

\section{Related Work}\label{rw}
In recent years, there has been an increase in online accounts advocating and supporting terrorist groups such as ISIS~\cite{TwitterPubliPolicy}. This phenomenon has attracted researchers to study their online existence, and research ways to automatically detect these accounts and limit their spread. Ashcroft et al.~\cite{Ashcroft2015} make an attempt to automatically detect Jihadist messages on Twitter. They adopt a machine-learning method to classify tweets as ISIS supporters or not. In the article, the authors focus on English tweets that contain a reference to a set of predefined English hashtags related to ISIS. Three different classes of features are used, including stylometric features, temporal features and sentiment features. However, one of the main limitations of their approach is that it is highly dependent on the data. Rowe and Saif~\cite{rowe2016} focused on studying Europe-based Twitter accounts in order to understand what happens before, during, and after they exhibit pro-ISIS behaviour. They define such behaviour as sharing of pro-ISIS content and/or using pro-ISIS terms. To achieve this, they use a term-based approach such that a user is considered to exhibit a radicalization behaviour if he/she uses more pro-ISIS terms than anti-ISIS terms. While such an approach seems effective in distinguishing radicalised users, it is unable to properly deal with lexical ambiguity (i.e., polysemy). Furthermore, in~\cite{kaati2015} the authors focused on detecting Twitter users who are involved with ``Media Mujahideen'', a Jihadist group who distribute propaganda content online. They used a machine learning approach using a combination of data-dependent and data-independent features. Similar to ~\cite{rowe2016} they used textual features as well as temporal features to classify tweets and accounts. The experiment was based on a limited set of Twitter accounts, which makes it difficult to generalize the results for a more complex and realistic scenario.

Radicalization literature also looked at psychological factors involved with adopting such behaviour. Torok~\cite{Torok2013} used a grounded theory approach to develop an explanatory model for the radicalization process utilizing concepts of psychiatric power. Their findings show that the process typically starts with the social isolation of individuals. This isolation seems to be self-imposed as individuals tend to spend a long time engaging with radical content. This leads to the concept of homophily, the tendency to interact and associate with similar others. Through constant interaction with like-minded people, an individual gradually strengthens their mindset and progresses to more extreme levels. Similarly, they start to feel as being part of a group with a strong group identity which leads to group polarization. In psychology, group polarization occurs when discussion leads the group to adopt actions that are more extreme than the initial actions of the individual group members~\cite{myers1976group}. Moreover, the National Police Service Agency of the Netherlands developed a model to describe the phases a Jihadist may pass through before committing an act of terrorism~\cite{elzinga2010}. These sequential phases of radicalism include strong links between the person's psychological and emotional state (e.g., social alienation, depression, lack of confidence in authority) and their susceptibility to radicalization.

\begin{figure*}[t]
    \centering
    \begin{subfigure}{.5\linewidth}
        \includegraphics[width=\columnwidth]{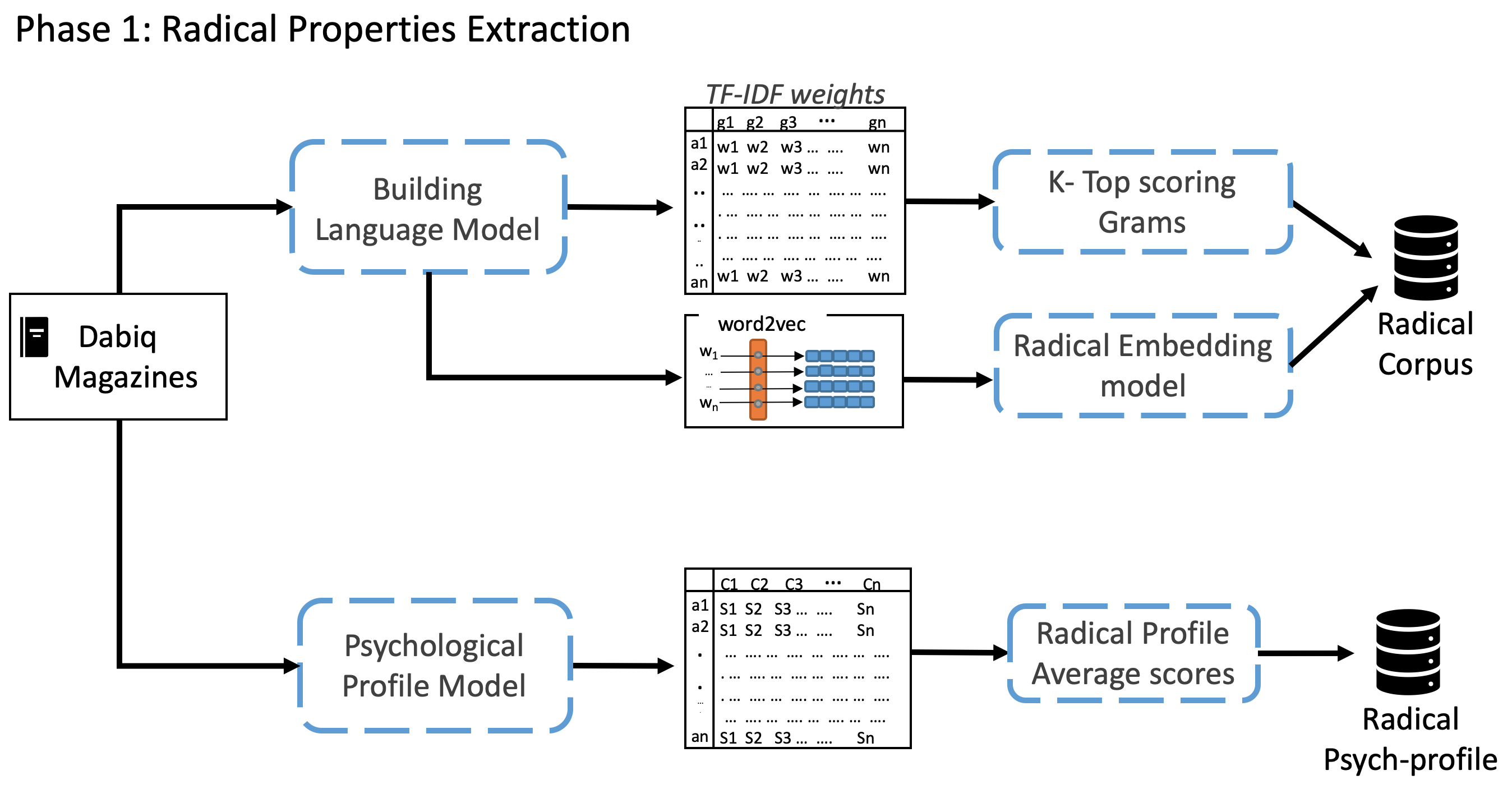}
    \end{subfigure}%
    ~~\vline ~~    
    \begin{subfigure}{.5\linewidth}
        \includegraphics[width=\columnwidth]{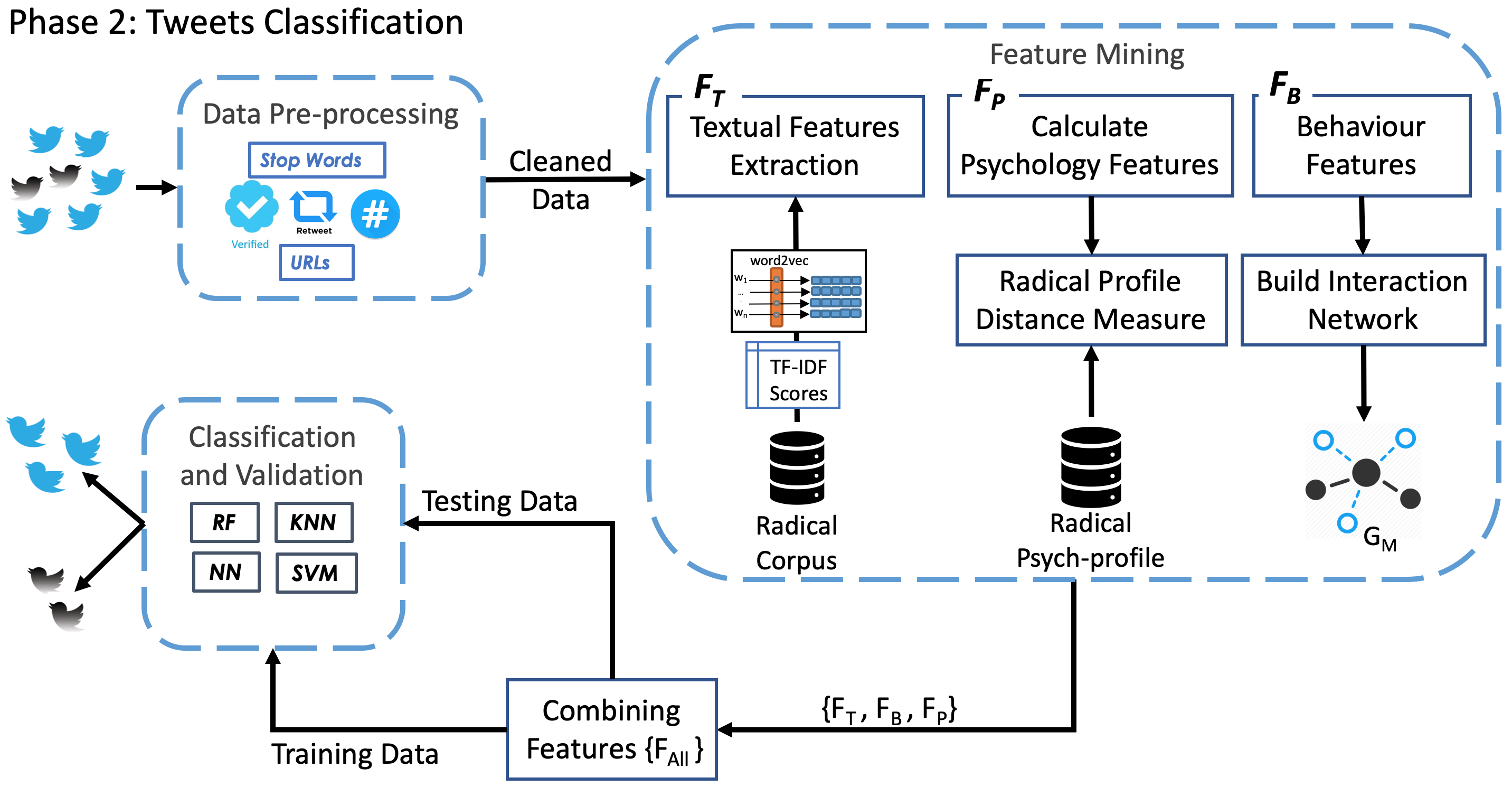}
    \end{subfigure}
    \caption{Approach overview}
    \label{fig:sysOver}
    \vspace{-10 pt}
\end{figure*}

\section{Methodology}\label{methods} 

As illustrated in Fig.~\ref{fig:sysOver}, our approach consists of two main phases: Phase 1:Radical Properties Extraction, where articles from Dabiq extremist magazines are input into this step to perform two parallel tasks. In the first task, we build a language model using (i) Term-Frequency Inverse-Document-Frequency (TF-IDF) scores of uni-, bi-, and tri-grams, and (ii) Word embeddings generated from a word2vec model~\cite{mikolov2013distributed}. The output of this task is a radical corpus of top k-grams, and a word embedding model giving a vector representation for each word in the corpus. The second task seeks to create a psychological profile based on the language used in the extremist propaganda articles, consisting of a set of emotional and topical categories using LIWC dictionary-based tool. Phase 2: Tweet classification involves the use of the models generated from Phase 1 to engineer features related to radical activities. We identify three groups of features and then train a binary classifier to detect radical tweets. 

\subsection{Feature Engineering}

Feature engineering is the process of exploring large spaces of heterogeneous features with the aim of discovering meaningful features that may aid in modeling the problem at hand. We explore three categories of information to identify relevant features to detect radical content. Some features are user-based while others are message-based. The three categories are: 1) Radical language (Textual features $F_T$);  2) Psychological signals (Psychological features $F_P$);  and 3) Behavioural features ($F_B$). In the following, we detail each of these categories.

\subsubsection{Radical Language}

In order to understand how radical messages are constructed and used, as mentioned earlier, we analyze content of ISIS propaganda material published in Dabiq magazine. Dabiq is an online magazine published by ISIS terrorist groups with the purpose of recruiting people and promoting their propaganda and ideology. Using this data source, we investigate what topics, textual properties, and linguistic cues exist in these magazines. Our intuition is that utilising these linguistic cues from the extremist propaganda would allow us to detect supporters of ISIS group who are influenced by their propaganda.

We use two methods to extract the radical language from the propaganda corpus. First we calculate tf-idf scores for each gram in the propaganda corpus. We use uni-grams, bi-grams, and tri-grams to capture phrases and context in which words are being used. We then select the top scoring grams to be used as features for the language model. N-grams and words frequency have been  used in the literature to classify similar problems, such as hate-speech and extremist text and have proven successful~\cite{Devyatkin2017}. The second method we use is word embeddings to capture semantic meanings. Research in NLP has compared the effectiveness of word embedding methods for encoding semantic meaning and found that semantic relationships between words
are best captured by word vectors within word embedding models~\cite{hamilton2016}. Therefore, we train word2vec model on our propaganda corpus to build the lexical semantic aspects of the text using vector space models. We learn word embeddings using skip-gram word2vec model implemented in the gensim package\footnote{\url{https://radimrehurek.com/gensim/models/word2vec.html}} with vector size of $100$ and window size of $5$. This word embedding model is used to obtain the vector representation for each word. We aggregate the vectors for each word in the tweet, and concatenate the maximum and average for each word vector dimension, such that any given tweet is represented in 200 dimension sized vector. This approach of aggregating vectors was used successfully in previous research~\cite{Liu2018ForecastingTP}. Moreover, since ISIS supporters typically advocate for violent behaviour and tend to use offensive curse words, we use dictionaries of violent words\footnote{\url{https://myvocabulary.com/word-list/terrorism-vocabulary}} and curse words\footnote{\url{https://www.cs.cmu.edu/~biglou/resources/bad-words.txt}} to record the ratio of such words in the tweet. We also count the frequency of words with all capital letters as they are traditionally used to convey yelling behaviour.

\subsubsection{Psychological Signals}
Research in fields such as linguistics, social science, and psychology suggest that the use of language and the word choices we make in our daily communication, can act as a powerful signal to detect our emotional and psychological states~\cite{liwc}. Several psychological properties are unintentionally transmitted when we communicate. Additionally, literature from the fields of terrorism and psychology suggests that terrorists may differ from non-terrorists in their psychological profiles~\cite{kruglanski2006psychology}. A number of studies looked at the motivating factors surrounding terrorism, radicalization, and recruitment tactics, and found that terrorist groups tend to target vulnerable individuals who have feelings of desperation and displaced aggression. In particular research into the recruiting tactics of ISIS groups, it was found that they focus on harnessing the individual's need for significance. They seek out vulnerable people and provide them with constant attention~\cite{pierson2017western}. Similarly, these groups create a dichotomy and promote the mentality of dividing the world into ``us'' versus ``them''~\cite{lopez9forensic}. Inspired by previous research, we extract psychological properties from the radical corpus in order to understand the personality, emotions, and the different psychological properties conveyed in these articles. 

We utilise LIWC dictionaries to assign a score to a set of psychological, personality, and emotional categories. Mainly, we look at the following properties:
\textbf{(1) Summary variables:} \textit{Analytically thinking} which reflects formal, logical, and hierarchical thinking (high value), versus informal, personal, and narrative thinking (low value). \textit{Clout} which reflects high expertise and confidence levels (high value), versus tentative, humble, and anxious levels (low value). \textit{Tone} which reflects positive emotions (high value) versus more negative emotions such as anxiety, sadness, or anger (low value). \textit{Authentic} which reflects whether the text is conveying honesty and disclosing (high value) versus more guarded, and distanced (low value).
\textbf{(2) Big five:} Measures the five psychological properties (OCEAN), namely Openness, Conscientiousness, Extraversion, Agreeableness, and Neuroticism.  
\textbf{(3) Emotional Analysis:} Measures the positive emotions conveyed in the text, and the negative emotions (including anger, sadness, anxiety).
\textbf{(4) Personal Drives:} Focuses on five personal drives, namely power, reward, risk, achievement, and affiliation. 
\textbf{(5) Personal Pronouns:} Counts the number of 1st, 2nd, and 3rd personal pronouns used. 
For each Twitter user, we calculate their psychological profiles across these categories. Additionally, using Minkowski distance measure, we calculate the distance between each of these profiles and the average values of the psychological properties created from the ISIS magazines. 

\subsubsection{Behaviour Signals}
This category consists of measuring behavioural features to capture different properties related to the user and their behaviour. This includes how active the user is (frequency of tweets posted) and the followers/following ratio. Additionally, we use features to capture users' interactions with others through using hashtags, and engagement in discussions using \textit{mention} action. To capture this, we construct the mention interaction graph ($G_M$) from our dataset, such that $G_M$ = $(U,E)$, where $U$ represents the user nodes and $E$ represents the set of edges. The graph $G_M$ is a directed graph, where an edge $e$ exists between two user nodes $A$ and $B$, if user $A$ mentions user $B$. After constructing the graph, we measure the degree of influence each user has over their network using different centrality measures, such as degree centrality, betweenness centrality, and HITS-Hub. Such properties have been adopted in the research literature to study properties of cyber-criminal networks and their behaviour~\cite{nouh2015identifying},~\cite{carrington2011crime}.

\section{Experiments}\label{exp}

\subsection{Dataset}~\label{data}
We acquired a publicly available dataset of tweets posted by known pro-ISIS Twitter accounts that was published during the 2015 Paris attacks by Kaggle data science community\footnote{\url{https://www.kaggle.com/fifthtribe/how-isis-uses-twitter/data}}. The dataset consists of around $17,000$ tweets posted by more than $100$ users. These tweets were labelled as being pro-ISIS by looking at specific indicators, such as a set of keywords used (in the user's name, description, tweet text), their network of follower/following of other known radical accounts, and sharing of images of the ISIS flag or some radical leaders. 
To validate that these accounts are indeed malicious, we checked the current status of the users' accounts in the dataset and found that most of them had been suspended by Twitter. This suggests that they did, in fact, possess a malicious behaviour that opposes the Twitter platform terms of use which caused them to be suspended. We filter out any tweets posted by existing active users and label this dataset as \textit{known-bad}. 

To model the normal behaviour, we collected a random sample of tweets from ten-trending topics in Twitter using the Twitter streaming API. These topics were related to news events and on-going social events (e.g., sports, music). We filter out any topics and keywords that may be connected to extremist views. This second dataset consists of around $8,000$ tweets published by around $1,000$ users. A random sample of $200$ tweets was manually verified to ascertain it did not contain radical views. We label this dataset as our~\textit{random-good} data.

A third dataset is used which was acquired from Kaggle community\footnote{\url{https://www.kaggle.com/activegalaxy/isis-related-tweets/home}}. This dataset is created to be a counterpoise to the pro-ISIS dataset (our known-bad) as it consists of tweets talking about topics concerning ISIS without being radical. It contains $122,000$ tweets from around $95,000$ users collected on two separate days. We verify that this dataset is indeed non radical by checking the status of users in Twitter and found that a subset ($24,000$ users) was suspended. We remove those from the dataset and only keep users that are still active on Twitter. This dataset is labelled as \textit{counterpoise} data. 

We performed a series of preprocessing steps to clean the complete dataset and prepare it for feature extraction. These steps are: (1) We remove any duplicates and re-tweets from the dataset in order to reduce noise. (2) We remove tweets that have been authored by verified users accounts, as they are typically accounts associated with known public figures. (3) All stop words (e.g., and, or, the) and punctuation marks are removed from the text of the tweet. (4) If the tweet text contains a URL, we record the existence of the URL in a new attribute, \textit{hasURL}, and then remove it from the tweet text. (5)  If the tweet text contains emojis (e.g., :-), :), :P), we record the existence of the emoji in a new attribute, \textit{hasEmj}, and then remove it from the tweet text. (6) If the tweet text contains any words with all capital characters, we record its existence in a new attribute, \textit{allCaps}, and then normalize the text to lower-case and filter out any non-alphabetic characters. (7)  We tokenize the cleansed tweet text into words, then we perform lemmatization, the process of reducing inflected words to their roots (lemma), and store the result in a vector.

\subsection{Experimental Set-up}

We conducted two experiments using the datasets described in Section~\ref{data}. Our hypothesis is that supporters of groups such as ISIS may exhibit similar textual and psychological properties when communicating in social media to the properties seen in the propaganda magazines. A tweet is considered radical if it promotes violence, racism, or supports violent behaviour. In \textit{Exp 1} we use the first two datasets, i.e., the \textit{known-bad} and the \textit{random-good} datasets to classify tweets to radical and normal classes. 
For \textit{Exp 2} we examine if our classifier can also distinguish between tweets that are discussing similar topics (ISIS related) by using the \textit{known-bad} and the \textit{counterpoise} datasets. 

The classification task is binomial (binary) classification where the output of the model predicts whether the input tweet is considered radical or normal. In order to handle the imbalanced class problem in the dataset, there are multiple techniques suggested in the literature Oversampling or undersampling of the minority/majority classes are common techniques. Another technique that is more related to the classification algorithm is cost sensitive learning, which penalizes the classification model for making a mistake on the minority class. This is achieved by applying a weighted cost on misclassifying of the minority class~\cite{chen2015using}. We will use the last approach to avoid downsampling of our dataset.

Previous research investigating similar problems reported better performances for Random Forest (RF) classifiers~\cite{Alfifi2017}. RF usually performs very well as it is scalable and is robust to outliers. RF typically outperforms decision trees as it has a hierarchical structure and is based on multiple trees. This allows RF to be able to model non-linear decision boundaries. Moreover, Neural Networks (NN) also produced good results when applied to problems related to image recognition, text and natural language processing~\cite{collobert2008unified}. However, they usually tend to require very large amounts of data to train. For the purpose of this study, we experimented with multiple classification algorithms, including RF, NN, SVM, and KNN and found that RF and NN produced the best performance. Due to space limitation, we only report results obtained using RF model. We configured the model to use $100$ estimators trees with a maximum depth of $50$, and we selected gini impurity for the split criteria. We used the out-of-bag samples (oob) score to estimate the generalization accuracy of the model. Additionally, since RF tends to be biased towards the majority class, we apply the cost sensitive learning method described earlier to make RF more suitable for imbalanced data~\cite{chen2015using}.

We divided the dataset to training set ($80$\%) and testing set ($20$\%), where the testing set is held out for validation. We reported validation results using different combinations of the features categories (i.e., $F_T$, $F_B$, $F_P$) and different evaluation metrics: accuracy, recall, precision, f-measure, and area under the ROC curve. Recall measures how many radical tweets we are able to detect, while precision measures how many radical tweets we can detect without falsely accusing anyone. For instance, if we identify every single tweet as radical, we will expose all radical tweets and thus obtain high recall, but at the same time, we will call everyone in the population a radical and thus obtain low precision. F-measure is the average of both precision and recall.

\subsection{Results}\label{result}

\begin{table}[t]
\centering
\scriptsize
\caption{Exp 1: Evaluation metrics across all feature groups}\label{exp1_results}
\begin{tabular}{|c|c|c|c|c|}
\hline      
Features        &  AC  & Precision & Recall    & F-measure \\ \hline 
$F_T(tf-idf)$   &  0.52  &    0.76   &  0.52        &  0.37 \\ \hline
$F_T(w2v)$      &  0.81  &   0.82    &    0.81      & 0.81  \\ \hline 
$F_T$           &  0.84  &    0.84  &   0.84      & 0.84  \\ \hline 
$F_B$           &  0.94  &    0.95  & 	0.94  	 	 & 0.94 \\ \hline 
$F_P$           &  1.0  &    1.0   	 &  1.0    	  	& 1.0 \\ \hline 
$F_{ALL}$       &  1.0  &     1.0   & 	1.0       	  & 1.0 \\ \hline 
\end{tabular}
\end{table}

\begin{table}[t]
\centering
\scriptsize
\caption{Exp 2: Evaluation metrics across all feature groups}\label{exp2_results}
\begin{tabular}{|c|c|c|c|c|}
\hline      
Features        &  AC   & Precision & Recall       & F-measure \\ \hline 
$F_T(tf-idf)$   &  0.56  &    0.69   &  0.56        &  0.48 \\ \hline
$F_T(w2v)$      &  0.73  &   0.73    &    0.73       & 0.73 \\ \hline 
$F_T$           &  0.80  &    0.80   &   0.80       & 0.80 \\ \hline 
$F_B$           &  0.91  &     0.92   & 0.91           & 0.91 \\ \hline 
$F_P$           &  1.0  &    1.0   &  1.0        & 1.0 \\ \hline 
$F_{ALL}$       &  1.0  &     1.0   & 1.0       & 1.0 \\ \hline 
\end{tabular}
\end{table}

\begin{figure}[t]
    \centering
    \begin{subfigure}{.85\linewidth}
        \centering
        \includegraphics[width=\columnwidth]{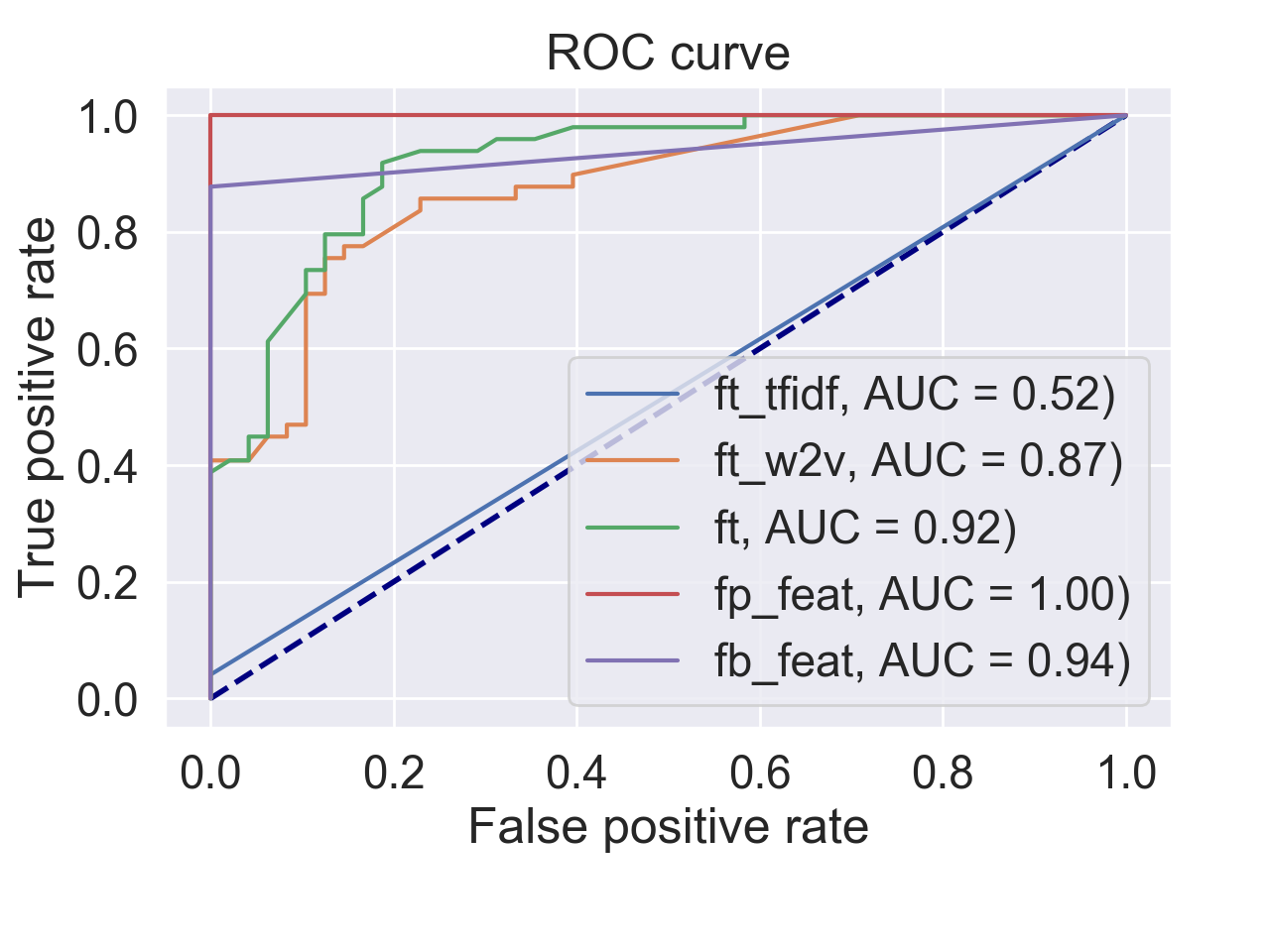}
    \end{subfigure}%
    \vspace{-10 pt}
    \begin{subfigure}{.85\linewidth}
        \centering
        \includegraphics[width=\columnwidth]{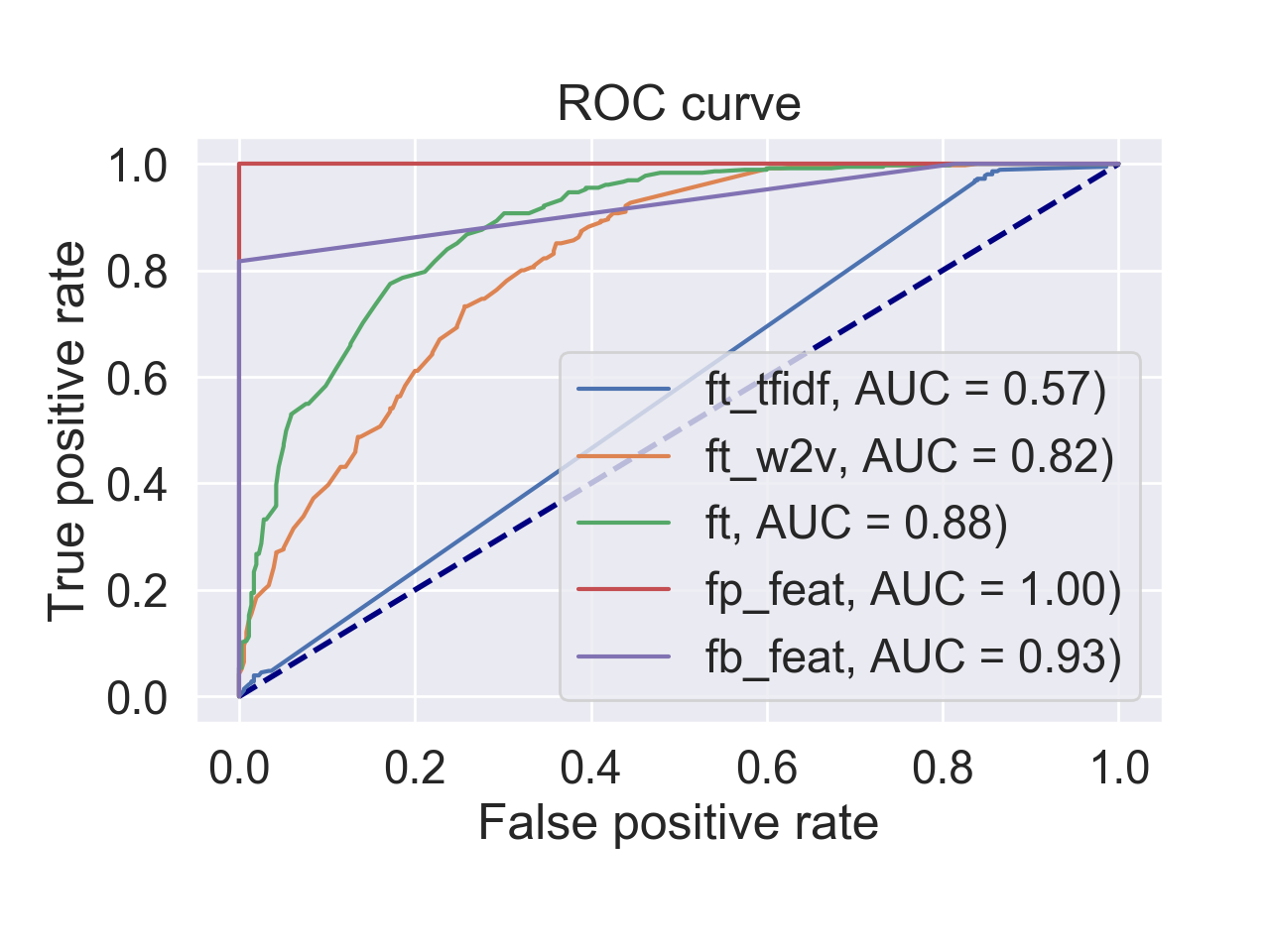}
    \end{subfigure}
    \vspace{-10 pt}
    \caption{ROC curve for Exp1 (top), Exp2 (bottom).}
    \label{fig:roc_curve_all}
    \vspace{-10 pt}
\end{figure}

\textbf{Exp 1:} The classification results using the \textit{known-bad} and \textit{random-good} datasets are reported in Table~\ref{exp1_results}. The table shows the average accuracy, precision, recall and f-measure scores obtained from each feature category ($F_{T}$, $F_{P}$, $F_{B}$) and their combination ($F_{All}$). We also compared the two textual models, and find that results obtained from using word embedding outperforms the use of n-grams tf-idf scores. This confirms that contextual information is important in detecting radicalization activities. Furthermore, our model performed best using the $F_{P}$ features across all metrics. This means that the model is able to distinguish between both radical and non-radical with high confidence using only $F_P$. 

\textbf{Exp2:} In this experiment, we tested the performance of our classifier in distinguishing between radical and normal tweets that discusses ISIS-related topics. Although this task is more challenging given the similarity of the topic discussed in the two classes, we find that the model still achieves high performance. Table~\ref{exp2_results} shows the different metrics obtained from each feature category. The $F_T$ feature group obtains $80$\% accuracy, and $91$\%, $100$\% for $F_B$ and $F_P$ feature groups, respectively. The results are consistent with the ones obtained from the first experiment with the features from $F_P$ group contributing to the high accuracy of the model. The area under the Receiver Operator Characteristic (ROC) curve, which measures accuracy based on \textit{TP}, and \textit{FP} rates, is shown in Fig.~\ref{fig:roc_curve_all} for each classification model.

\subsection{Features Significance}
We investigated which features contribute most to the classification task to distinguish between radical and non-radical tweets. We used the mean decrease impurity method of random forests~\cite{louppe2013understanding} to identify the most important features in each feature category. The ten most important features are shown in Table~\ref{table-imp}. We found that the most important feature for distinguishing radical tweets is the psychological feature distance measure. This measures how similar the Twitter user is to the average psychological profile calculated from the propaganda magazine articles. Following this is the Us-them dichotomy which looks at the total number of pronouns used (I,they, we, you). This finding is in line with the tactics reported in the radicalization literature with regards to emphasizing the separation between the radical group and the world.  

Moreover, among the top contributing features are behavioural features related to the number of mentions a single user makes, and their HITS hub and authority rank among their interaction network. This relates to how active the user is in interacting with other users and how much attention they receive from their community. This links to the objectives of those radical users in spreading their ideologies and reaching out to potential like-minded people. As for the $F_T$ category, we find that the use of word2vec embedding improves the performance in comparison with using the tf-idf features. Additionally, all bi-grams and tri-grams features did not contribute much to the classification; only uni-grams did. This can be related to the differences in the writing styles when constructing sentences and phrases in articles and in the social media context (especially given the limitation of the number of words allowed by the Twitter platform). Additionally, the \textit{violent word ratio}, \textit{longWords}, and \textit{allCaps} features are among the top contributing features from this category. This finding agrees to a large extent with observations from the literature regarding dealing with similar problems, where the use of dictionaries of violent words aids with the prediction of violent extremist narrative.

\begin{table}[t]
\centering
\scriptsize
\caption{Features Importance}
\label{table-imp}
\begin{tabular}{|l|l|c|}
\hline      
Top 10          &  	Features  	&  Category	\\ \hline 
1				&	Radical psych-profile distance	&	$F_P$\\ \hline
2				&	Us-Them	dichotomy			   &	$F_P$\\ \hline
3				&	\# of mentions a user make	   &	$F_B$ 	\\ \hline
4				&	User rank (hub and authority)	&	$F_B$\\ \hline
5				&	Sad emotion						&	$F_P$	\\ \hline
6				&	Risk driver					&	$F_P$ 	\\ \hline
7				&	All-caps count			&	$F_T$ 	\\ \hline
8				&	URL count				&	$F_T$ \\ \hline
9				&	Violent-word ratio		&	$F_T$ \\ \hline
10				&	Hash count				&	$F_T$ \\ \hline

\end{tabular}
\vspace{-10 pt}
\end{table}

\section{Conclusion and Future Work}\label{conclustion}
In this paper, we identified different signals that can be utilized to detect evidence of online radicalization. We derived linguistic and psychological properties from propaganda published by ISIS for recruitment purposes. We utilize these properties to detect pro-ISIS tweets that are influenced by their ideology. Unlike previous efforts, these properties do not only focus on lexical keyword analysis of the messages, but also add a contextual and psychological dimension. We validated our approach in different experiments and the results show that this method is robust across multiple datasets. 
This system can aid law enforcement and OSN companies to better address such threats and help solve a challenging real-world problem. In future work, we aim to investigate if the model is resilient to different evasion techniques that users may adopt. We will also expand the analysis to other languages.

\bibliographystyle{IEEEtran} 
\bibliography{ref}

\end{document}